\documentclass[11pt,a4paper]{article}
\usepackage{jcappub}
\usepackage{silence} 
\usepackage[english]{babel}
\usepackage[utf8]{inputenc}
\usepackage[OT2,T1]{fontenc}
\usepackage{graphicx}
\usepackage{orcidlink} 
\usepackage{amsmath,amssymb,amsfonts,amsthm}
\usepackage{bm,bbm}
\usepackage{tikz,stackrel}
\usepackage{pstricks}
\usepackage{pifont} 
\usepackage{empheq}
\newcommand\nn{\nonumber\\}

\newcommand{\bc}{\begin{center}}
	\newcommand{\ec}{\end{center}}
\newcommand{\be}{\begin{equation}}
	\newcommand{\ee}{\end{equation}}
\newcommand{\ba}{\begin{eqnarray}}
	\newcommand{\ea}{\end{eqnarray}}
\def\bs{\begin{subequations}}
	\def\es{\end{subequations}}
\newcommand{\ben}{\begin{equation*}}
	\newcommand{\een}{\end{equation*}}
\newcommand{\ban}{\begin{eqnarray*}}
	\newcommand{\ean}{\end{eqnarray*}}
\renewcommand{\leq}{\leqslant}

\def\b{\beta}

\def\Om{\Omega}

\def\s{\sigma}

\def\cT{\mathcal{T}}

\newcommand{\Eq}[1]{(\ref{#1})}

\newcommand{\Eqqs}[1]{eqs.~(\ref{#1})}

\def\cob{\color{blue}}

\newcommand{\au}[2]{#1.~#2}
\newcommand{\book}[5]{\emph{#1}, #2, #3, #4 (#5)}
\newcommand{\books}[4]{\emph{#1}, #2, #3 (#4)}
\newcommand{\oarX}[1]{\href{http://arxiv.org/abs/#1}{{\ttfamily\cob arXiv:#1}}}
\newcommand{\arX}[1]{\href{http://arxiv.org/abs/#1}{{\ttfamily\cob arXiv:#1}}}
\newcommand{\doin}[6]{\href{http://dx.doi.org/#1}{{\cob {\it #2} {\bf #3 #4} (#6) #5}}}
\newcommand{\doinn}[5]{\href{http://dx.doi.org/#1}{{\cob {\it #2} {\bf #3} (#5) #4}}}
\newcommand{\doij}[5]{\href{http://dx.doi.org/#1}{{\cob {\it #2} {\bf #3} (#5) #4}}}

\newcommand{\tia}[1]{\textit{#1},}

\def\Mpl{M_{\rm Pl}}

\def\rme{e}
\def\rmd{d}

\newcounter{listcounter}

\allowdisplaybreaks

\usepackage{float}
\usepackage{comment}
\usepackage{enumitem}
\usepackage{subcaption}
\usepackage[normalem]{ulem}

\def\laq{~\raise 0.4ex\hbox{$<$}\kern -0.8em\lower 0.62ex\hbox{$\sim$}~}
\def\gaq{~\raise 0.4ex\hbox{$>$}\kern -0.7em\lower 0.62ex\hbox{$\sim$}~}

\def\beq{\begin{equation}}
	\def\eeq{\end{equation}}
\def\bea{\begin{eqnarray}}
	\def\eea{\end{eqnarray}}

\def \b {\beta}

\def \Om {\Omega}

\begin{document}
	
\renewcommand{\thefootnote}{\fnsymbol{footnote}}
	
\title{LISA as a probe of pre-big-bang physics: a nested sampling analysis}

\author[a]{Xo\'an Vilas Curr\'as\,\orcidlink{0009-0001-6649-4301},}
\emailAdd{xoan.cosmo@gmail.com}
\affiliation[a]{Departamento de Astrofísica, Universidad de La Laguna, Avenida Astrofísico Francisco Sánchez, 38200 Tenerife, Spain}
	
\author[b,*]{Gianluca Calcagni\,\orcidlink{0000-0003-2631-4588}\note{Corresponding author.}}
\emailAdd{g.calcagni@csic.es}
\affiliation[b]{Instituto de Estructura de la Materia, CSIC, Serrano 121, 28006 Madrid, Spain}

\abstract{Using a nested sampling analysis, we study the gravitational-wave background (GWB) predicted by the Gasperini--Veneziano model of pre-big-bang cosmology, both in its most recent minimal and non-minimal versions. Within the LISA sensitivity range, the GWB signal is a flat or a broken power law, parametrized by four fundamental quantities: the Hubble parameter at the curvature bounce $H_1$, the axion mass $m$, the initial amplitude of the axion field $\sigma_i$ and the exponent $\beta$ governing the high-energy growth of the dilaton and the dynamics of the internal dimensions. We determine the posterior distributions of these parameters based on how LISA would detect such signal. Including the galactic and extra-galactic foregrounds in the analysis, the most stringent constraints on $H_1$, $\sigma_i$ and $\beta$ are obtained when the signal exhibits a left-bend feature, while for $m$ this happens for a right-bend feature. Relative uncertainties reach $\Delta H_1/H_1 ,\,\Delta m/m \sim 18\%$ at $68\%$ confidence level under favourable conditions. LISA will thus be capable of placing significant constraints on the pre-big-bang model and, in the event of detection, providing some preliminary empirical hints of low-energy string-inspired dynamics.}


\maketitle
\renewcommand{\thefootnote}{\arabic{footnote}}


\section{Introduction}

Gravitational waves (GWs) are a recent discovery \cite{Abbott:2016blz} that is rapidly populating our previously blind view of the cosmos with sources \cite{LIGOScientific:2025slb,LIGOScientific:2025pvj} and revolutionizing our approach to astrophysics and cosmology. Each advance in GW detection has so far confirmed the validity of Einstein's classical theory of general relativity (GR) \cite{TheLIGOScientific:2016src,LIGOScientific:2021sio,LIGOScientific:2025cmm,LIGOScientific:2025rid}. However, attempting to describe our observable universe entirely with a classical theory may lead to an incomplete description or explanation of phenomena where quantum mechanics plays a major role. GWs offer a unique opportunity to test not only GR but also its modifications beyond the classical level. In particular, primordial GWs were generated before photon decoupling and can provide valuable information about epochs prior to the formation of the cosmic microwave background, since unlike light they interact very weakly with matter and thus travel almost unimpeded across the universe.

GWs can be generated in two main ways: by astrophysical sources (galactic and extra-galactic black-hole, neutron-star and white-dwarf binaries) or by cosmological sources (inflation, phase transitions, cosmic strings and primordial black holes) \cite{Maggiore:2019uih,LISA:2022yao,LISACosmologyWorkingGroup:2022jok,LISA:2022kgy,Colpi:2024xhw,Abac:2025saz}. In the first case, GWs can arrive either from individual sources or as a stochastic superposition of many signals forming gravitational-wave background (GWB) \cite{Christensen:2018iqi,Renzini:2022alw,vanRemortel:2022fkb}. In the second case, GWs manifest themselves primarily as a GWB.

Among the cosmological mechanisms, one case of particular interest is the GWs produced around the big bang itself. This event marking the beginning of Everything arises from extrapolating GR to the origin of time. However, at epochs close to $t = 0$ the classical theory breaks down and quantum effects become as important as, or even dominate over, the classical gravitational force.
Therefore, to properly describe the origin of the universe, a theory of quantum gravity might be required. Among many other candidates capable of incorporating quantum mechanics and gravity \cite{Ori09,Fousp,Bam24}, string theory has been employed to address this issue. And, among many other models of string or string-inspired cosmology, one in particular has the potential of leaving an imprint in GW physics: the Gasperini--Veneziano pre-big-bang model of the early universe \cite{Gasperini:1992em,Gasperini:2002bn,Gasperini:2007vw,Gasperini:2007zz,Gasperini:2016gre,Gasperini:2021mat,Gasperini:2023tus,Conzinu:2023fth,Ben-Dayan:2024aec,Tan:2024urn,Conzinu:2024cwl,Tan:2024qgk,Conzinu:2025sot}. While not derived from a complete formulation of string theory, this scenario provides a distinctive phenomenological setting in which string-motivated ingredients can leave an observable imprint in the GW sector.

According to this paradigm, a consequence of T-duality \cite{Veneziano:1991ek} is that the big bang does not represent the true origin of time but, rather, a moment of maximum curvature and energy density, implying the existence of a preceding phase from $t =-\infty$ to $t = 0$, known as the pre–big-bang era. If this model is correct, then GWs can be generated during the pre–big-bang phase and leave a relic in the form of a cosmological GWB. The existence of a GWB has already been confirmed by observations in a relatively narrow frequency range \cite{NANOGrav:2023gor,NANOGrav:2023hvm,EPTA:2023fyk,Reardon:2023gzh,Xu:2023wog,InternationalPulsarTimingArray:2023mzf} but the current challenge lies not only in scrutinizing its possible astrophysical origin and identifying individual contributions from different sources, but also in spanning a larger frequency range in search for a continuation of this signal or for new ones of different origin. Accomplishing this highly demanding task requires high-resolution measurements across various frequency bands, which are expected to be provided by future GW detectors such as the Laser Interferometer Space Antenna (LISA) \cite{LISA:2022yao,LISACosmologyWorkingGroup:2022jok,LISA:2022kgy,Colpi:2024xhw} and the Einstein Telescope \cite{Maggiore:2019uih,Abac:2025saz}. The bulk of the signal of the pre-big-bang GWB should fall within the sensitivity frequency window of these two observatories in particular \cite{Ben-Dayan:2024aec}. In this publication, the GWB of the model was plotted against the sensitivity curves of LISA and Einstein Telescope but little was said about the constraints these detectors will be able to place upon the pre-big-bang scenario.

Until LISA and the Einstein Telescope become operational in the 2030s, we must rely on numerical simulations to ``test'' theoretical models. In this work, we build upon the framework developed in \cite{Conzinu:2024cwl} to recast the pre-big-bang GWB in terms of the parameter space made of physical constants. Then, we naturally continue the analysis of \cite{Ben-Dayan:2024aec} and apply a LISA-specialized \texttt{Python} package, \texttt{SGWBinner} \cite{Caprini:2019pxz,Pieroni:2020rob,Flauger:2020qyi}, to compute posterior distributions for various test cases, thus preparing the ground for future detection efforts with LISA.

The conclusions of this work indicate that LISA has the potential to place significant constraints on the model. The latter would emerge as a compelling high-energy candidate, although it would not constitute a confirmation of the model since other, more conventional early-universe scenarios can produce a similar signal. In this case, additional and complementary multi-messenger observations would be needed to disentangle this degeneracy. Conversely, if LISA set bounds incompatible with the theoretically viable parameter space, the model in its present form would be ruled out. A third and, for obvious reasons, less attractive possibility is that the parameter space were constrained to a region where the signal is undetectable by LISA, thus relegating the model to a minor spectator to be ``further investigated in the future.''

This paper is organized into the following sections: a brief review of the pre-big-bang cosmology model in section~\ref{sec:pbb_cosmology}, a description of the parameter space used in our analysis in section~\ref{sec:param_space}, the methodology and results obtained using \texttt{SGWBinner} in section~\ref{sec:Results} and their discussion in section~\ref{sec:Discusion}. Conclusions are presented in section~\ref{sec:conclusions}. The derivation of some formul\ae\ can be found in the appendices. We work in natural units $\hbar = 1=c$ and $\Mpl=1$.


\section{Pre-big-bang cosmology}\label{sec:pbb_cosmology}

The pre-big-bang model originally proposed by Gasperini and Veneziano \cite{Gasperini:1992em} and later refined by them and other authors \cite{Gasperini:2002bn,Gasperini:2007vw,Gasperini:2007zz,Gasperini:2016gre,Gasperini:2021mat,Gasperini:2023tus,Conzinu:2023fth,Ben-Dayan:2024aec,Tan:2024urn,Conzinu:2024cwl,Tan:2024qgk,Conzinu:2025sot} is an attempt to embed a description of the early Universe  using elements of string theory. One starts from the 10-dimensional low-energy superstring action $S[\phi,H_{ABC}, g_{AB}]$, where $\phi$ is the dilaton, $H_{ABC}$ is the Kalb--Ramond field strength and $g_{AB}$ is the 10-dimensional metric of target spacetime. It is assumed that geometry can be factorized into a four-dimensional ``external'' spacetime (a flat Friedmann--Lema\^itre--Robertson--Walker background plus perturbations) and an ``internal'' six-dimensional manifold:
\be
\rmd s^2 = a^2(\tau)\left(-\rmd\tau^2+\rmd\bm{x}^2\right)+\sum_{i=1}^6 b_i^2(\tau)\,\rmd y_i^2\,,
\ee
where $\tau$ is conformal time. It is also assumed that all dynamical variables only depend on the external coordinates \cite{Gasperini:2002bn,Gasperini:2016gre}. The dilaton $\phi$, non-minimally coupled to gravity in the Jordan frame, plays the role of the inflaton in driving the cosmic accelerated expansion, while the Kalb--Ramond axion sources the seeds of scalar fluctuations. The big bang singularity is replaced by a state of maximal curvature and energy density. In the pre-bounce phase, the Universe starts in the perturbative string vacuum with zero curvature and string coupling $g_{\rm s}$; quantum perturbations trigger the dilaton and axion dynamics and the curvature begins to grow with the dilaton, until it reaches a maximum. After this stage, the internal dimensions and the dilaton field become stabilized
through some presumably non-perturbative mechanism, whose detailed realization is still not completely understood. The Kalb--Ramond axion decays, leading to a conventional reheating phase. This string cosmological dynamics is based upon several gravitational equations containing higher-curvature and quantum string-loop corrections, in principle at all orders. Explicit examples of background solutions describing a regular bounce transition can be found in \cite{Gasperini:2023tus,Conzinu:2023fth} (regular bounce at all orders in $\alpha'$), \cite{Gasperini:2004ss} (which includes non-local effects) and \cite{Gasperini:2007zz}.

Before proceeding, we mention some limitations of the model which are still under debate. On one hand, both the Calabi--Yau space and modulus stabilization are assumed without an explicit construction. While an agnostic attitude towards the structure of the internal space could be regarded as flexible, it is not clear whether the pre-big-bang model is naturally realized in the string landscape. Also, moduli stabilization can occur by string-loop corrections or non-perturbative potentials but these possibilities have not been inspected in full in this model. Currently, the explicit realization of a smooth bounce and of the stabilization of the moduli and the dilaton is available only in specific examples \cite{Gasperini:2007zz,Gasperini:2023tus,Conzinu:2023fth}. On the other hand, these open questions are related to a third one, namely, the problem of initial conditions and the justification to start with a quasi-homogeneous four-dimensional spacetime \cite{Gasperini:1999bn}. Large homogeneous regions in the Jordan frame may arise from the gravitational collapse of gravity and dilatonic waves in the initial state of the universe \cite{Buonanno:1998bi} but this remains a hypothesis.
All of this suggests taking any empirical finding on the pre-big-bang model with caution, although it should not deter one to analyze the model with the best tools available.

We outline here the derivation of the GWB of this model, omitting various intermediate steps, as a full review falls outside the scope of this work. For a comprehensive treatment, the reader is referred to \cite{Ben-Dayan:2024aec,Conzinu:2024cwl,Gasperini:2007zz}. As detailed therein, the spectral shape $\Om_{\textsc{gw}}$ of the GWB  is
\begin{equation}\label{eq:omega_original}
    \Om_{\textsc{gw}} = \frac{1}{\rho_{\rm crit}(\tau_0)}\frac{\rmd\rho_k(\tau_0)}{\rmd\ln k}\,,
\end{equation}
where $\tau_0$ denotes the current value of conformal time $\tau$, $\rho_{\text{crit}} = 3 M_{\text{Pl}}^2 H^2$ defines the critical energy density and $M_{\text{Pl}}^2 = (8\pi G)^{-1}$ is the reduced Planck mass. The quantity $\rho_k(\tau_0)$ is the energy density of the $k$-th Fourier mode of tensor perturbations, which are amplified by the specific early-universe scenario under consideration and evaluated at $\tau_0$. We can write the differential energy of the amplified perturbation as 
\begin{equation}\label{eq:drho}
    \rmd\rho_k(\tau_0) = 2k\langle n_k(\tau_0)\rangle\frac{\rmd^3 \textit{\textbf{k}}}{(2 \pi)^3} = \frac{k^4}{\pi^2}\langle n_k(\tau_0) \rangle \rmd\ln k\,,
\end{equation}
 where $\langle n_k(\tau_0) \rangle$ is the number density of gravitons produced at $\tau_0$. This quantity is determined by solving the evolution equation for the Fourier mode of the canonical Mukhanov--Sasaki variable $\upsilon_k=\xi_h h_k$ obeying
\begin{equation}\label{eq:evo_pumpf}
    \upsilon_k'' + \left( k^2 - \frac{\xi_h''}{\xi_h} \right)\upsilon_k = 0\,,
\end{equation}
where a prime denotes differentiation with respect to the conformal time and \cite{Gasperini:2016gre}
\be
\xi_h=a\,\rme^{-\frac\phi2}\prod_{i=1}^6 b_i^\frac12\,.
\ee
In particular, in the post-bounce regime the pump field coincides with the scale factor in the Einstein frame. The scalar perturbations sourced by the axion obey a similar equation but with a different background function (or ``pump field'') $\xi_\s=a^2\xi_h^{-1}$. The pump fields $\xi_h(\tau)$ and $\xi_\s(\tau)$ represent the way the evolving cosmological background ``pumps'' energy into perturbation modes. Their time dependence encodes the effects of the scale factor $a(\tau)$ and of the dilaton $\phi(\tau)$, determining when and how quantum fluctuations are amplified.

The pre-big-bang model can follow two main scenarios: a \emph{minimal} one where the scalar spectrum does not change slope and S-duality is preserved \cite{Ben-Dayan:2024aec} and a \emph{non-minimal} one where either or both conditions (monotonic spectrum and S-duality) are broken \cite{Conzinu:2024cwl,Conzinu:2025sot}. Both scenarios yield the same parametric solutions within each cosmological phase and, consequently, produce the same form of GWB. The primary distinction lies in the extension of the parameter space of the model, which is discussed in detail in the following sections.

Following the analyses in \cite{Ben-Dayan:2024aec,Conzinu:2024cwl}, the cosmic evolution is characterized by five different phases:
\begin{enumerate}
\item Low-energy string: $\xi_h \propto (-\tau)^{1/2}, \qquad \tau < -\tau_s$.
\item de Sitter-like evolution: $\xi_h \propto \left( -\tau \right)^{\beta-1}, \qquad -\tau_s < \tau < -\tau_1$.
\item Radiation domination: $\xi_h \propto \tau,\qquad -\tau_1 < \tau < \tau_\sigma$.
\item Matter domination: $\xi_h \propto \tau^2, \qquad \tau_\sigma < \tau < \tau_d$.
\item Radiation domination: $\xi_h \propto \tau,\qquad \tau > \tau_d$.
\end{enumerate}
Here $-\tau_s$ marks the transition from a low-energy initial phase to a possible late-time attractor; $-\tau_1$ marks the time scale of the curvature bounce; $\tau_\sigma$ marks the beginning of a dust-like phase dominated by axion oscillations; and $\tau_d$ marks the decay of the axion, associated with conventional reheating. The parameter $\beta$ describes the high-energy growth of the dilaton and the dynamics of the internal dimensions.

\begin{figure}[H]
    \centering
    \includegraphics[width=\linewidth]{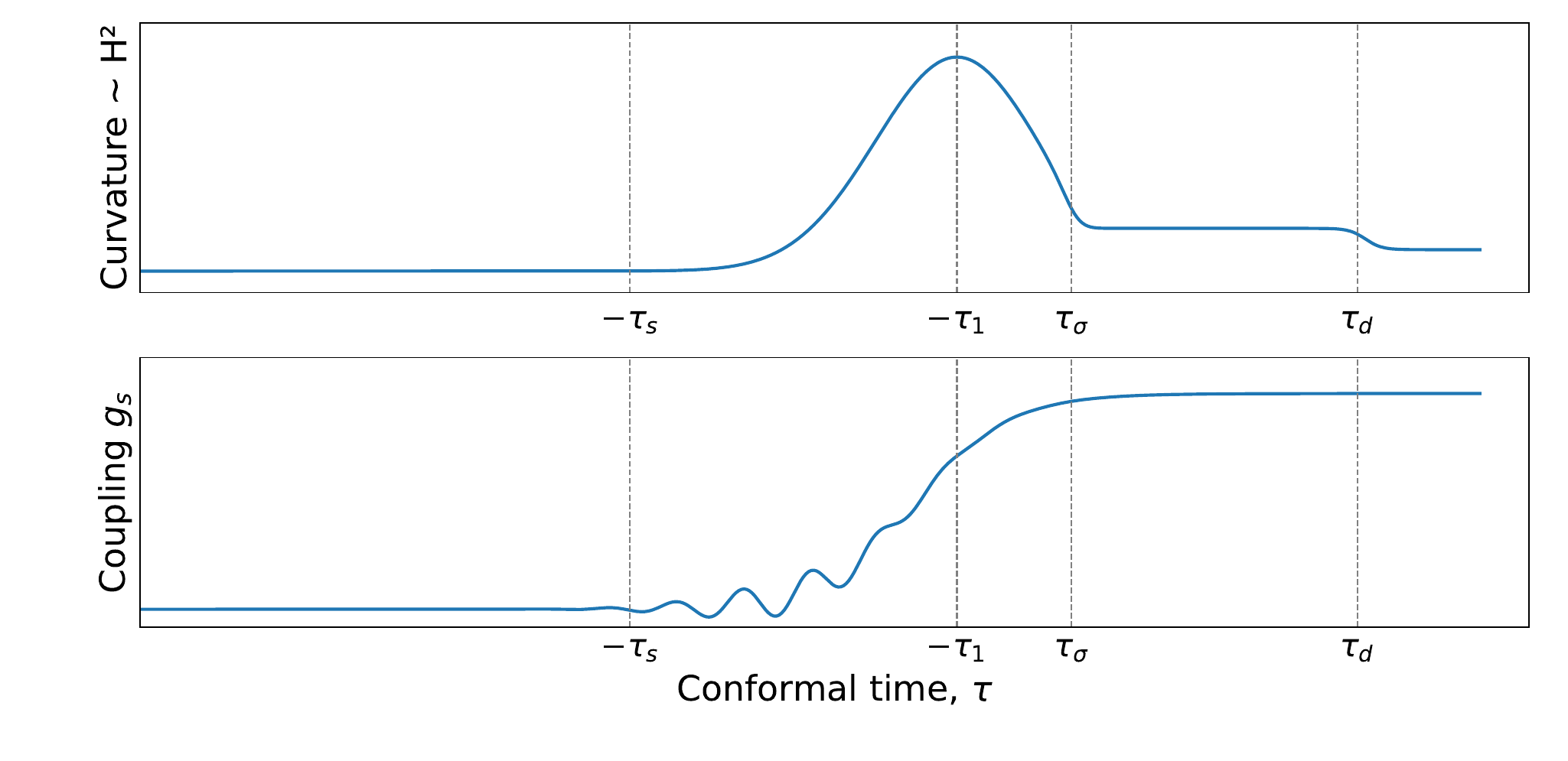}
    \caption{Schematic evolution of the curvature scale $H^2$ and of the string coupling $g_s = e^{\phi/2}$ in the pre–big/bang scenario. The universe starts in a low–curvature, weakly coupled regime in the asymptotic past, followed by a phase of growing curvature and dilaton. The high–curvature regime is reached around the string scale, after which the background evolves into the standard post–big-bang radiation- and matter-dominated phases. The figure is meant to provide a qualitative illustration of the background dynamics and does not represent a specific numerical solution.}
    \label{fig:curvature_coupling}
\end{figure}

The qualitative behaviour of the curvature scale and of the string coupling across these phases is schematically illustrated in Fig.~\ref{fig:curvature_coupling}.

Let us now turn to perturbations. Each conformal time scale can be conveniently replaced by its corresponding curvature scale, denoted by \textbf{$H_s$}, \textbf{$H_1$}, \textbf{$H_\sigma$} and \textbf{$H_d$}. It is convenient to define the wave-number $k$ associated with primordial GW modes, focusing on the interval $k_0 \leq k \leq k_1$, where $k_0 \equiv \tau_0^{-1}$ is the minimal wave-number corresponding to the largest observable scale today and $k_1 \equiv \tau_1^{-1}$ denotes the maximal frequency that exited the horizon at the end of the pre–big-bang inflationary phase, meaning that they are inside the present Hubble horizon. This wave-number can be converted into the corresponding proper frequency $f$ via the relation $\omega(\tau) = k/a(\tau)$, where $a(\tau)$ is the scale factor at conformal time $\tau$ and $\omega = 2\pi f$. Lastly, the following dimensionless parameters are often used in the literature:
\begin{equation}\label{eq:def_zetas}
    z_s \coloneqq \frac{\tau_s}{\tau_1} = \frac{k_1}{k_s}\,, \qquad z_\sigma \coloneqq \frac{\tau_\sigma}{\tau_1}=\frac{k_1}{k_\sigma}\,, \qquad z_d \coloneqq \frac{\tau_d}{\tau_1} = \frac{k_1}{k_d}\,,
\end{equation}
which control the extension of the pre-bouncing high curvature regime and of the two post-bouncing phases.

Equation \Eq{eq:evo_pumpf} can be solved noting that the pump fields $\xi_{h,\s}$ follow a simple power law during each cosmic phase:
\be
\xi_h \sim (-\tau)^{-1+\beta_h}\,,\qquad \xi_\sigma \sim  (-\tau)^{-1+\beta_\sigma} \,,\qquad \beta_\sigma = - \beta_h + \epsilon\,.
\ee
The minimal case is recovered when $\epsilon = 0$.

To compute the spectral distribution of the GWB, one solves \Eq{eq:omega_original} using \Eq{eq:drho}. This yields the amplitude of the signal as a function of the number density of produced gravitons. As an illustrative example, consider the frequency range between $k_\sigma$ and $k_1$:
\ben
\Omega_{\textsc{gw}}(k,\tau) = \frac{\omega^4}{\pi^2\rho_{\rm crit}(\tau)} \langle n_\omega(\tau) \rangle ,
\qquad k_\sigma < k < k_1\, .
\een
By rewriting the critical density and the graviton occupation number as
\ben
\rho_{\rm crit}(\tau) = \frac{\rho_r(\tau)}{\Omega_r(\tau)}\, ,
\qquad
\langle n_\omega(\tau) \rangle \simeq \left( \frac{k}{k_1} \right)^{-1-\lvert 3-2\beta \rvert} ,
\een
one obtains 
\ben
\Omega_{\textsc{gw}}(k,\tau) =
\frac{\Omega_r(\tau)}{\Omega_{r0}}
\Omega_{\rm PBB}
\left( \frac{k}{k_1} \right)^{3-\lvert 3-2\beta \rvert} ,
\qquad
k_\sigma < k < k_1 , \qquad \tau > \tau_d\,.
\een
The same procedure can be extended to the remaining frequency ranges, allowing us to reconstruct the full gravitational-wave spectrum. In terms of the characteristic frequencies $f_1$, $f_\s$, $f_d$ and $f_s$, the GWB generated in each phase takes the piece-wise asymptotic form
\begin{equation}\label{eq:spl}
\renewcommand{\arraystretch}{2}
    \Om_{\textsc{gw}}(f) = 
    \left\{
    \begin{array}{ll}
    \Omega_{\mathrm{PBB}} \left( \displaystyle\frac{f}{f_1} \right)^{3-|3-2\beta|}, &\quad f_\sigma \lesssim f \lesssim f_1 \\
    \Om_{\textsc{gw}}(f_1) \left( \displaystyle\frac{f_\sigma}{f_1} \right)^{3-|3-2\beta|} \left( \displaystyle\frac{f}{f_\sigma} \right)^{1-|3-2\beta|}, &\quad f_d \lesssim f \lesssim f_\sigma \\
    \Om_{\textsc{gw}}\left(f_\sigma\right) \left( \displaystyle\frac{f_d}{f_\sigma} \right)^{1-|3-2\beta|} \left( \displaystyle\frac{f}{f_d} \right)^{3-|3-2\beta|}, &\quad f_s \lesssim f \lesssim f_d \\
    \Om_{\textsc{gw}}\left(f_d\right) \left( \displaystyle\frac{f_s}{f_d} \right)^{3-|3-2\beta|} \left( \displaystyle\frac{f}{f_s} \right)^3, &\quad \hspace{0.9cm}f \lesssim f_s
    \end{array}
    \right.
\end{equation}
where $\Omega_{\mathrm{PBB}}$ is the following normalization constant dependent on the radiation density $\Omega_{\mathrm{r}}$:
\be
    \Omega_{\mathrm{PBB}}\coloneqq\Omega_{\mathrm{r0}} H_1^2 \left( \frac{z_\sigma}{z_d} \right)^2 \,, \qquad\Omega_{\mathrm{r0}} \equiv \Omega_{\mathrm{r}}(\tau_0) \approx 4.15 \times10^{-5} h^{-2}\,.
\ee
This derivation follows closely the treatment presented in \cite{Conzinu:2024cwl}, to which the reader is referred for a more detailed and comprehensive discussion.

The range of interest for LISA is the third from the top in \Eq{eq:spl}, $f_s \lesssim f \lesssim f_d$ \cite{Ben-Dayan:2024aec}. A case of special interest is $\beta = 0$, for which the signal is maximized to a plateau. Since $\Om_{\textsc{gw}}(f)\propto f^{3-|3-2\beta|} = f^{3-|3-0|} = f^0 = 1$, there is no frequency dependence within the observation range for the chosen value of $\beta$. As a result, the piecewise function based on simple power laws is not applicable in this scenario and the complete expression must be used instead \cite{Ben-Dayan:2024aec}:
\ba
\Om_{\textsc{gw}}(f) &=& \,\Omega_{\mathrm{PBB}}f^3(f^2+f_s^2)^{-\frac{|3-2\beta|}{2}}(f^2+f_d^2)^{-1}(f^2+f_\sigma^2)(f^2+f_1^2)^{\frac{|3-2\beta|-3}{2}} \nn
        &&\times \exp\left(-\frac{f}{f_1} + \arctan \frac{f}{f_1}\right).\label{eq:bpl}
\ea
Reducing this equation to the regime of interest ($f_1 \gg f$, $f_\sigma \gg f$), we get the broken power law
\be\label{eq:mi_modelo}
    \Om_{\textsc{gw}}(f)=\alpha \frac{f^3}{(f^2+f_s^2)^{\frac{|3-2\beta|}{2}}(f^2+f_d^2)}\,,\qquad f_s \lesssim f \lesssim f_d\,,
\ee
where $\alpha = \Omega_{\mathrm{PBB}} f_\sigma^2 f_1^{|3-2\beta|-3}$.

The above characteristic frequencies (in Hz) are functions of the fundamental parameters of the model (appendix \ref{appA}):
\be\label{eq:def_freqs}
    \scalebox{1.1}{$
    \begin{array}{ll}
    &f_1 \simeq  \dfrac{3.9\times10^{11}}{2\pi}H_1^{1/2}m^{1/3}\sigma_i^{-2/3} \,,\qquad f_\sigma \simeq \dfrac{3.9\times10^{11}} {2\pi}m^{5/6}\sigma_i^{4/3} \,,\\
    &f_d \simeq \dfrac{3.9\times10^{11}}{2\pi}m^{3/2} \,,\qquad  f_s = \dfrac{3.9\times10^{11}}{2\pi z_s}H_1^{1/2}m^{1/3}\sigma_i^{-2/3} \,,
    \end{array}
    $}
\ee
where the parameters $z_s$, $z_\s$ and $z_d$ defined in \Eq{eq:def_zetas} are (appendix \ref{appA})
\ba
\log_{10}z_s &=&  \frac{1}{1-n_{\rm s}-2\beta} \left[ \frac{5-n_{\rm s}}{2} \log_{10}H_1  - \log_{10}\left( \frac{4.2\pi^2}{\mathcal{T}^2(\sigma_i)} \right) \right.\nn
&&\left.+ 9 - (1-n_{\rm s})(\log_{10}1.5 - 27)  - \frac{n_{\rm s}-1}{2}\log_{10}\left(m^{2/3}\sigma_i^{-4/3}\right) \right] ,\label{z_s}\\
z_\sigma &=& H_1^{1/2} m^{-1/2}\sigma_i^{-2}\,,\qquad z_d = H_1^{1/2} m^{-7/6} \sigma_i ^{-2/3}.
\ea
Here $n_{\rm s}$ is the spectral index of primordial scalar perturbations and $\cT$ is the transfer function \cite{Ben-Dayan:2024aec}. Formula \Eq{z_s} follows from the fact that axion perturbations seed curvature perturbations through the curvaton mechanism, in order to match the observed cosmic-microwave-background anisotropies \cite{Ben-Dayan:2024aec}. From the above expressions, $\Omega_{\mathrm{PBB}}$ can be rewritten as
\begin{equation}
    \Omega_{\mathrm{PBB}} = \Omega_{\mathrm{r0}} H_1^2 m^{4/3}\sigma_i^{-8/3}\,.
\end{equation}

\begin{figure}[H]
    \centering
    \includegraphics[height=0.75\textwidth, width=1\linewidth]{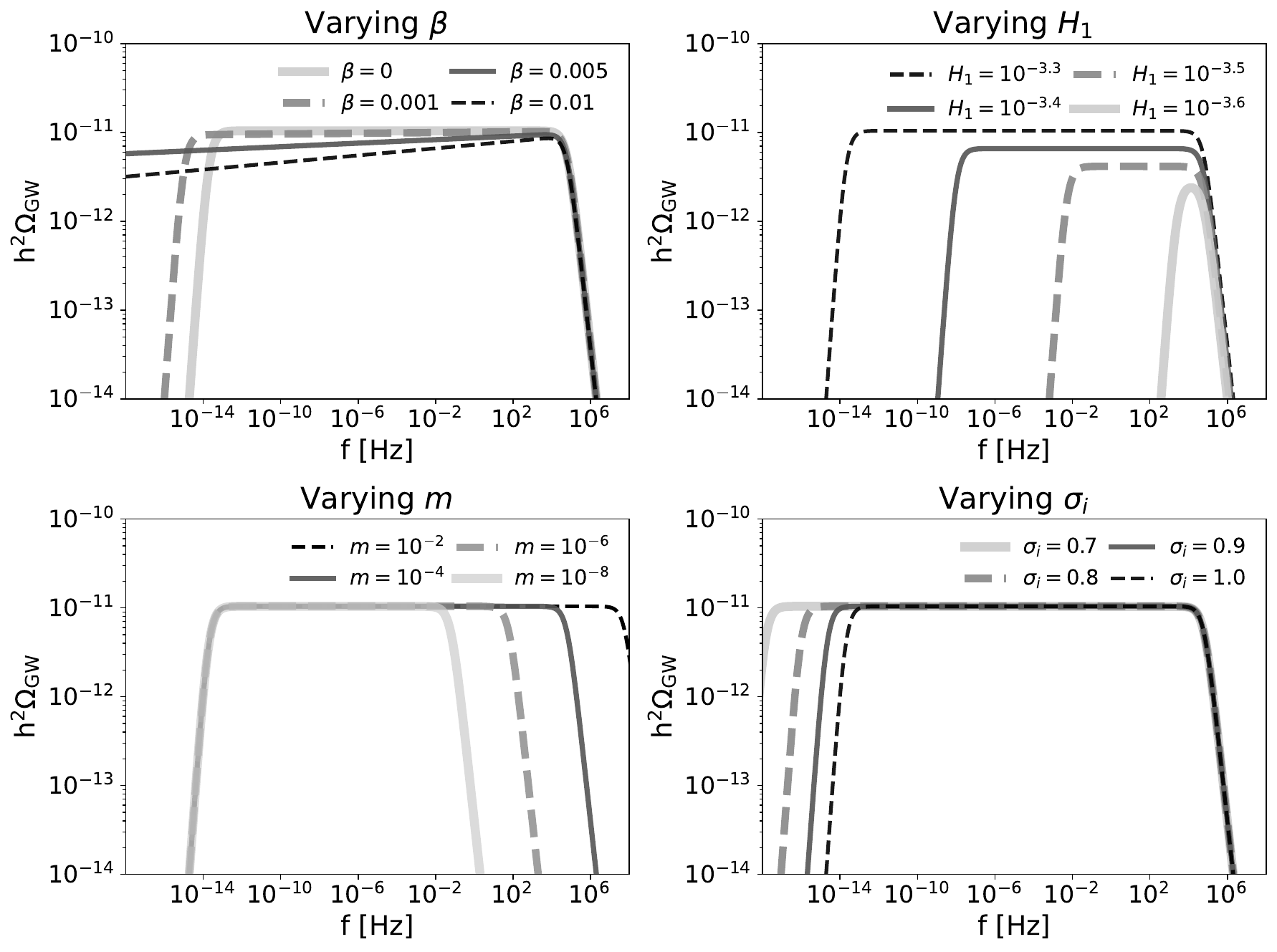}
    \caption{Dependence of GWB test curves on the parameters $\b$, $H_1$, $m$ and $\s_i$.}
    \label{fig:variacion_parametros}
\end{figure}

In this work, we parametrize $\Om_{\textsc{gw}}(f)$ \Eq{eq:mi_modelo} in terms of $\beta$ (which describes the high-energy growth of the dilaton and the dynamics of the internal dimensions; it was denoted as $\beta_h$ in \cite{Conzinu:2024cwl}), $H_1$ (the Hubble parameter when the curvature bounces), $m$ (the axion mass) and $\sigma_i$ (the initial axion amplitude after the bounce). This parametrization is chosen for two reasons. First, it is based on the fundamental parameters of the theory and it thus obviates the problem of parameter reconstruction in GW physics, i.e., how to pass from the posterior distributions of the phenomenological parameters used in numerical simulations to the theoretical ones. This differs from a template-based approach such as the one followed in \cite{LISACosmologyWorkingGroup:2024hsc}, where one can test several models simultaneously using generic templates but at the price of facing the reconstruction problem. Second, it allows us to catalogue the effect of each parameter on the GWB curve. For visualization purposes only, a representation of how the variation of each parameter influences the curve is provided in figure~\ref{fig:variacion_parametros}.

Each parameter affects the signal as follows.
\begin{itemize}
    \item As the value of $\beta$ moves further away from 0 to positive values, the left bend of the curve shifts to lower frequencies and the amplitude decreases.
    \item The smaller the value of $H_1$, the more the left bend shifts to higher frequencies and the amplitude decreases.
    \item The smaller the value of $m$, the more the right bend shifts to lower frequencies.
    \item The smaller the value of \textbf{$\sigma_i$}, the more the left bend shifts to lower frequencies.
\end{itemize}


\section{Parameter space}\label{sec:param_space}


As mentioned earlier, the only difference between the minimal and non-minimal scenarios lies in the boundary conditions imposed. Both of them share the following conditions:
\begin{itemize}
    \item Growing string coupling and avoidance of background instabilities. In the minimal model,
\begin{equation}
    0\leq \beta < 3\,.
\end{equation}
    \item Hierarchy of the transition frequency scales:
\begin{equation}\label{eq:limitesf}
f_s<f_d<f_\sigma\lesssim f_1\,.
\end{equation}
    \item The axion background field must be oscillating when it becomes dominant:
\begin{equation}\label{eq:limites1}
    \log_{10}H_1 + \frac{3}{2}\log_{10}z_d - \frac{7}{2}\log_{10}z_\sigma \leq 0\,.
\end{equation}
    \item The end of the axion dynamics must happen before the big-bang nucleosynthesis scale:
\begin{equation}\label{eq:limites2}
    \log_{10}H_1 - 3 \log_{10} z_d + \log_{10} z_\sigma > -42 - \log_{10}4\,.
\end{equation}
\end{itemize}
A detailed derivation of these inequalities can be found in \cite{Ben-Dayan:2024aec,Conzinu:2024cwl}. These can be inverted and expressed in terms of the parameters used in this work, yielding the following boundary conditions (appendix~\ref{appB}):
\be\label{eq:limites}
0 < \beta \leq 3 \,, \qquad 10^{-42.6} < H_1 < 1\,, \qquad
10^{-14.2} < m < 1\,, \qquad 10^{-7.1} < \sigma_i < 1\,.
\ee

As explained in section \ref{sec:pbb_cosmology}, an extension of the parameter space in the $\beta<0$ direction is theoretically possible \cite{Conzinu:2024cwl}. 
 This possibility was used to reach the signal detected by the International Pulsar Timing Array (IPTA), which requires $\Om_{\textsc{gw}}(f_s) \approx 2.9^{+5.4}_{-2.3}\times10^{-8}$, $f_s \approx (1.2 \pm 0.6) \times 10^{-8} \,\mathrm{Hz}$. In turn, these values are achieved for $10^{-3} \lesssim H_1 \lesssim 10^{-1}$, $10^{-3/2} \lesssim \sigma_i \lesssim 1$ (both within the range \Eq{eq:limites}) and negative values of $\beta$ typically of order $|\b|=O(10^{-3})$ but up to as low values as $-0.064$ \cite{Conzinu:2024cwl}. The resulting spectrum corresponds to an extreme case, characterized by large amplitude gradients across frequencies.

In the computational part of the present work, we slightly extend the parameter space of the minimal model to include also negative values of $\b$ admitted by the non-minimal model. In particular, we set the lower bound for the prior on $\beta$ to $-0.05$, well above the observational constraint $\b>-0.19$ \cite{Tan:2024qgk}. Compared to the $|\b|=O(10^{-3})$ values explored in \cite{Conzinu:2024cwl}, we consider this to be a fairly conservative starting point to explore the parameter space, ensuring that both the minimal and the non-minimal scenario are treated on equal footing. Moreover, all the values of $H_1$ we explore are significantly larger than $10^{-10}$, which is compatible with the fact that $H_1$ represents the bouncing scale and must be of order of the string mass, independently of the specific scenario considered. We verified that reducing the lower bound for $H_1$ to $10^{-10}$ yields identical results to those obtained by using the full range in \Eq{eq:limites}, while improving computational efficiency. On the other hand, we do not explore all the parameter space of the non-minimal model. For instance, we keep $H_1$ away from 1 and consider injected signals with a medium-to-low amplitude, thus avoiding very-large signals that would drown any other source within the LISA window and produce extreme amplitude gradients towards lower frequencies. 

Consequently, the final parameter-space boundaries adopted in this work consist of a moderate extension of \Eq{eq:limites} to the non-minimal scenario, a shrinking of the range of $H_1$ and an innocuous rounding of the priors on $m$ and $\s_i$:
\be\label{eq:boundaries}
-0.05 < \beta \leq 3\,,\qquad 10^{-10} < H_1 < 1\,,\qquad
10^{-14} < m < 1\,,\qquad 10^{-7} < \sigma_i < 1\,.
\ee
The density of these priors is uniform for $\b$ and $\s_i$ (in the latter case, this is feasible because we inject values at or above $0.2$) and log-uniform for $H_1$ and $m$.


\section{Methodology and results}\label{sec:Results}

With the theoretical form of the signal (\ref{eq:mi_modelo}) and the parameter validity ranges (\ref{eq:boundaries}), we employed the \texttt{Python} package \texttt{SGWBinner} \cite{Caprini:2019pxz,Pieroni:2020rob,Flauger:2020qyi} to determine the posterior distribution of the fundamental parameters of the model given an injected signal and the LISA specifications. We included the noise, the galactic and the extra-galactic foregrounds, letting $d=8$ free parameters: four theoretical, two of the noise (low-frequency and high-frequency amplitude) and two of the foregrounds (amplitudes).

We adopted the nested sampling method, in particular, the PolyChord algorithm \cite{Handley:2015vkr}. For cross-checking, we also used the Markov Chain Monte Carlo (MCMC) method \cite{Lewis:2002ah,Lewis:2013hha}, which is implemented within the Cobaya framework \cite{Torrado:2020dgo,Cobaya2}. In this work, we only present the results obtained with the first method.

PolyChord uses a cloud of points that is gradually reduced in number as the algorithm converges. It is controlled by three key parameters, the first being the number of live points ($n_{\rm live}$), which determines the initial size of the point cloud; then, the algorithm calculates the likelihood of each point and converts the ones with the lowest likelihood into dead points. For each dead point, a new live point with higher likelihood is generated by slice sampling, each slice sampling chain having $n_{\rm repeats}$ steps. The procedure continues until the uncertainty in the evidence, calculated from both live and dead points, reaches a desired threshold (precision criterion). The parameter configuration used in this work is as follows: number of live points $80d$ (where $d=8$ is the number of free parameters), number of repeats $2d$ and a precision criterion of $10^{-5}$. We checked that the results are robust if we reduce the number of live points to $60d$ while increasing the number of repeats to $3d$.

\begin{figure}[ht!]
    \centering
    
    \begin{subfigure}[b]{0.45\textwidth}
        \includegraphics[width=\linewidth]{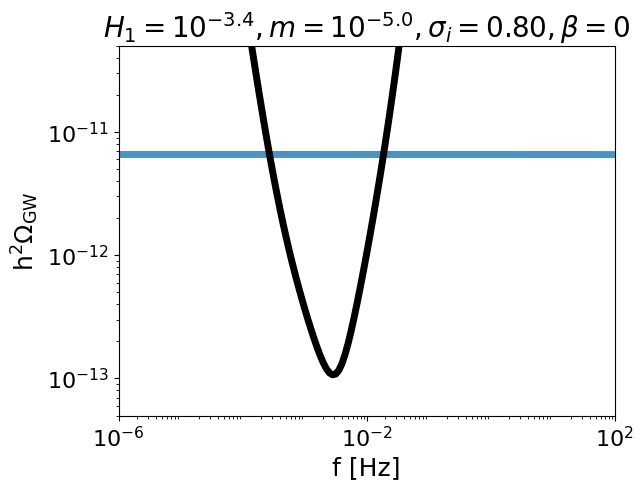}
        \caption{high-SNR plateau}
        \label{fig:highsnr}
    \end{subfigure}
    \hfill
    \begin{subfigure}[b]{0.45\textwidth}
        \includegraphics[width=\linewidth]{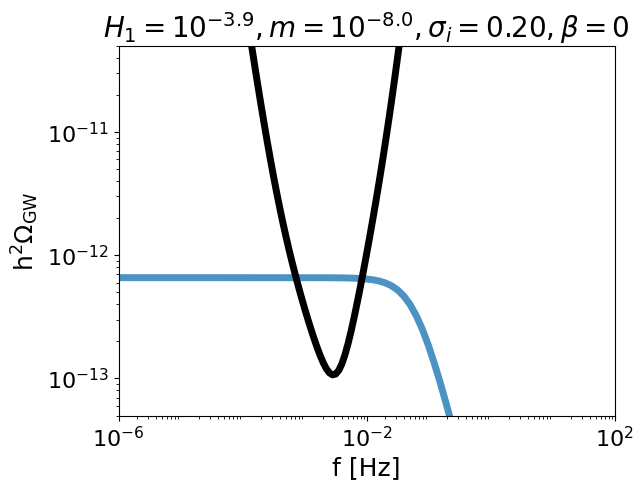}
        \caption{low-SNR plateau}
        \label{fig:lowsnr}
    \end{subfigure}
    
    \vspace{0.3cm}
    
    \begin{subfigure}[b]{0.45\textwidth}
        \includegraphics[width=\linewidth]{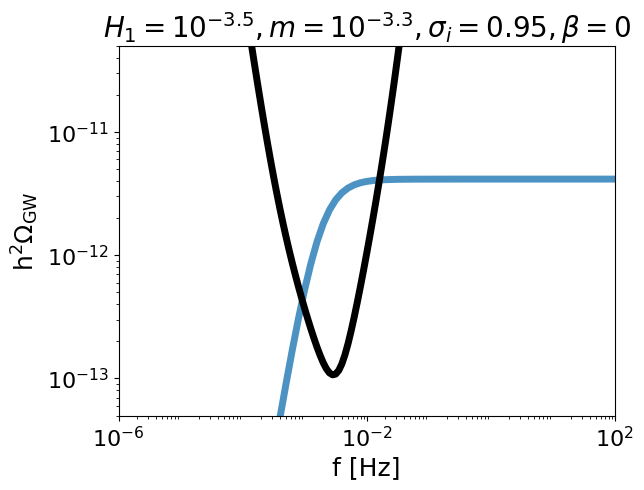}
        \caption{left bend}
        \label{fig:leftbend}
    \end{subfigure}
    \hfill
    \begin{subfigure}[b]{0.45\textwidth}
        \includegraphics[width=\linewidth]{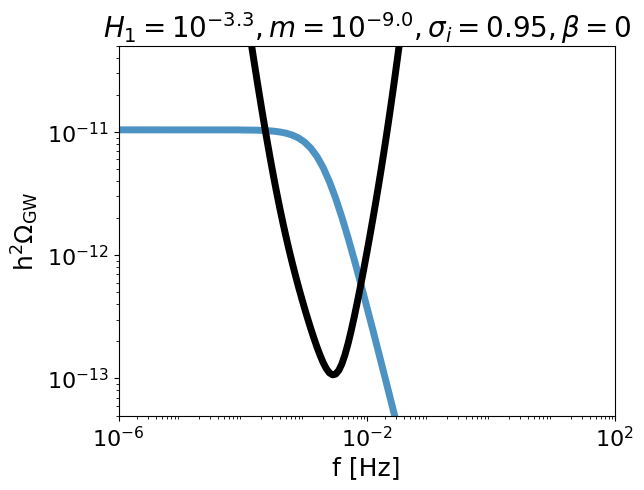}
        \caption{right bend}
        \label{fig:rightbend}
    \end{subfigure}
    
    \caption{The four test curves of the injected signals: a plateau with high SNR (top left); a plateau with low SNR (top right); a left-bend broken power law (bottom left); a right-bend broken power law (bottom right). The LISA sensitivity curve is shown in black and the injected signal in light blue.}
    \label{fig:snr-shapes}
\end{figure}

Regarding the injected signals, we chose four typical cases (figure \ref{fig:snr-shapes}): left-bend feature, right-bend feature and a plateau with high or low signal-to-noise ratio (SNR). We also considered a fifth case with a high-SNR plateau with negative $\b=-0.004$ but its results are very similar to the analogous $\b=0$ case. A sixth case, namely of no detection (that can be realized by injecting any signal lying outside the LISA window), is discussed in section~\ref{sec:conclusions}.


Once the posterior distribution for each parameter in each case has been obtained, two types of information can be extracted: their bounds and their relative uncertainties. Relative uncertainties are
\begin{equation}
    \textrm{Relative uncertainty on $A$:} \qquad \frac{\Delta A}{A_c} = \frac{A_{\mathrm{max}}-A_{\mathrm{min}}}{A_c}\,,
\end{equation}
where $A_c\neq 0$ is the injected value of the parameter $A$ and $A_{\mathrm{max}}$ and $A_{\mathrm{min}}$ are the highest and lowest value at a certain confidence level. For parameters such as $\b$ where $A_c=0$, the difference $A_{\mathrm{max}}-A_{\mathrm{min}}$ is considered instead.

\begin{figure}[ht]
    \centering
    \begin{tikzpicture}
        \node[inner sep=0pt] (img) at (0,0) {\includegraphics[width=\textwidth]{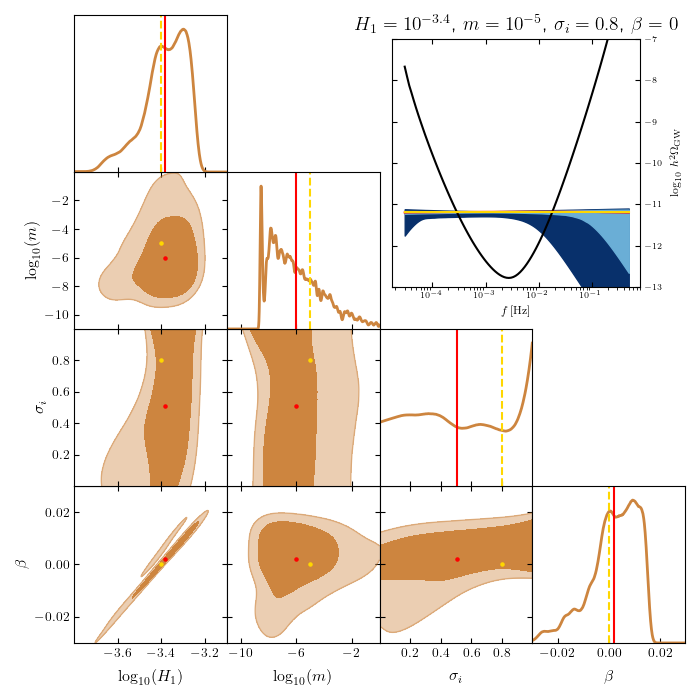}};
        \node[anchor=north west, text=black] at ([xshift=10.26cm,yshift=0.23cm]img.north west) {\fontsize{12.7}{12}\selectfont Injected signal};
    \end{tikzpicture}
    \caption{Top right: injected test signal of the pre-big-bang scenario (yellow curve) and reconstructed signal (red curve), compared with the LISA sensitivity curve (solid black); shaded regions in light (dark) blue correspond to the 68\% (respectively, 95\%) CL uncertainty in the reconstructed signal. Triangle plot: Two- and one-dimensional posterior probability distributions of the test-curve parameters. In the marginalized plot, the true and reconstructed value are marked by, respectively, a dashed yellow and a solid red line. The injected values are $\beta = 0$, $H_1 = 10^{-3.4}$, $m = 10^{-5.0}$ and $\sigma_i = 0.80$. Galactic and extra-galactic foregrounds are included.}
    \label{fig:curva_-3.4_gl}
\end{figure}

\begin{figure}[ht]
    \centering
    \begin{tikzpicture}
        \node[inner sep=0pt] (img) at (0,0) {\includegraphics[width=\textwidth]{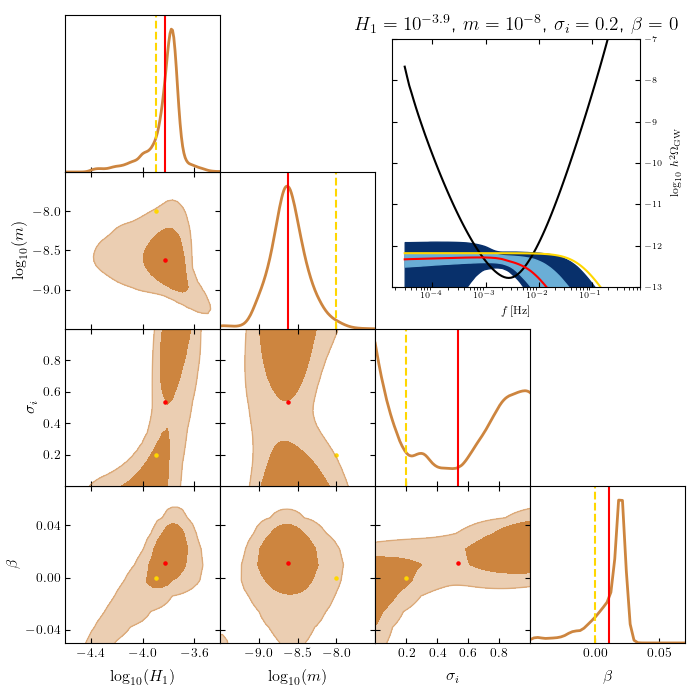}};
        \node[anchor=north west, text=black] at ([xshift=10.26cm,yshift=0.23cm]img.north west) {\fontsize{12.7}{12}\selectfont Injected signal};
    \end{tikzpicture}
    \caption{Top right: injected test signal of the minimal pre-big-bang scenario (yellow curve) and reconstructed signal (red curve), compared with the LISA sensitivity curve (solid black); shaded regions in light (dark) blue correspond to the 68\% (respectively, 95\%) CL uncertainty in the reconstructed signal. Triangle plot: Two- and one-dimensional posterior probability distributions of the test-curve parameters. In the marginalized plots, the true and reconstructed values are marked by, respectively, a dashed yellow and a solid red line. The injected values are $\beta = 0$, $H_1 = 10^{-3.9}$, $m = 10^{-8.0}$ and $\sigma_i = 0.20$. Galactic and extra-galactic foregrounds are included.}
    \label{fig:curva_-3.9_gl}
\end{figure}

\begin{figure}[ht]
    \centering
    \begin{tikzpicture}
        \node[inner sep=0pt] (img) at (0,0) {\includegraphics[width=\textwidth]{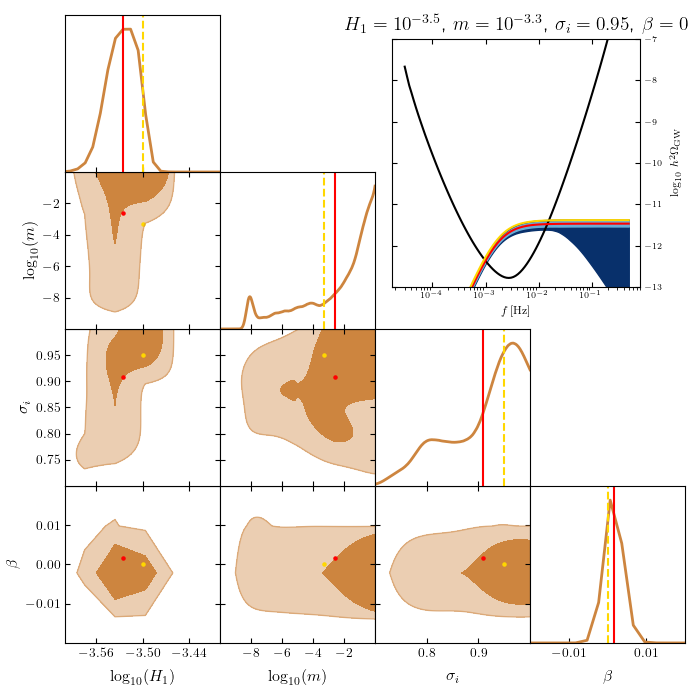}};
        \node[anchor=north west, text=black] at ([xshift=10.26cm,yshift=0.23cm]img.north west) {\fontsize{12.7}{12}\selectfont Injected signal};
    \end{tikzpicture}
    \caption{Top right: injected test signal of the minimal pre-big-bang scenario (yellow curve) and reconstructed signal (red curve), compared with the LISA sensitivity curve (solid black); shaded regions in light (dark) blue correspond to the 68\% (respectively, 95\%) CL uncertainty in the reconstructed signal. Triangle plot: Two- and one-dimensional posterior probability distributions of the test-curve parameters. In the marginalized plots, the true and reconstructed values are marked by, respectively, a dashed yellow and a solid red line. The injected values are $\beta = 0$, $H_1 = 10^{-3.5}$, $m = 10^{-3.3}$ and $\sigma_i = 0.95$. Galactic and extra-galactic foregrounds are included.}
    \label{fig:curva_-3.5_gl}
\end{figure}

\begin{figure}[ht]
    \centering
    \begin{tikzpicture}
        \node[inner sep=0pt] (img) at (0,0) {\includegraphics[width=\textwidth]{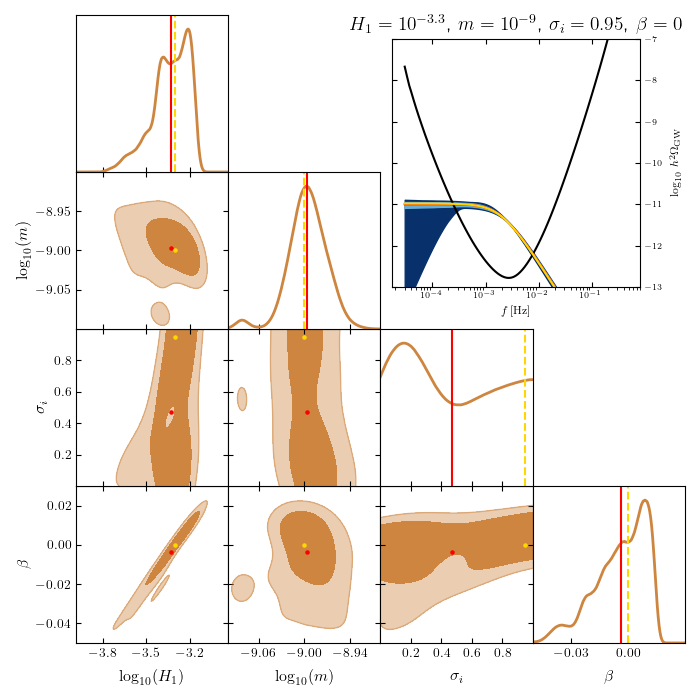}};
        \node[anchor=north west, text=black] at ([xshift=10.26cm,yshift=0.23cm]img.north west) {\fontsize{12.7}{12}\selectfont Injected signal};
    \end{tikzpicture}
    \caption{Top right: injected test signal of the minimal pre-big-bang scenario (yellow curve) and reconstructed signal (red curve), compared with the LISA sensitivity curve (solid black); shaded regions in light (dark) blue correspond to the 68\% (respectively, 95\%) CL uncertainty in the reconstructed signal. Triangle plot: Two- and one-dimensional posterior probability distributions of the test-curve parameters. In the marginalized plots, the true and reconstructed values are marked by, respectively, a dashed yellow and a solid red line. The injected values are $\beta = 0$, $H_1 = 10^{-3.3}$, $m = 10^{-9.0}$ and $\sigma_i = 0.95$. Galactic and extra-galactic foregrounds are included.}
    \label{fig:curva_-3.3_gl}
\end{figure}

The results of the nested sampling analysis performed on the injected signals presented in figure~\ref{fig:snr-shapes} are shown in figures~\ref{fig:curva_-3.4_gl}--\ref{fig:curva_-3.3_gl}. Both the galactic and the extra-galactic foregrounds are included.

\newpage

\newpage

\newpage

\clearpage


\section{Discussion}\label{sec:Discusion}


Let us discuss the bounds and relative uncertainty for each parameter shown in figures~\ref{fig:curva_-3.4_gl}--\ref{fig:curva_-3.3_gl}.
\begin{itemize}
    \item $\bm{\beta}$: The lowest precision in the estimation of $\beta$ occurs for a right-bend configuration, while the highest precision is achieved for the left-bend feature (figure~\ref{fig:curva_-3.5_gl}). The 95\% CL intervals for $\beta$ can be summarized as follows: $|\beta|< 0.011$ in the case of a left bend (figure~\ref{fig:curva_-3.5_gl}); $|\beta|<0.022$ for a plateau with high SNR (figure~\ref{fig:curva_-3.4_gl}); $|\beta|< 0.045$ for a plateau with low SNR (figure~\ref{fig:curva_-3.9_gl}); $-0.04<\beta< 0.02$ in the case of a right bend (figure~\ref{fig:curva_-3.3_gl}). This behaviour is consistent with the fact that variations in $\beta$ induce a slope on the low-frequency side of the curve, as illustrated in figure~\ref{fig:variacion_parametros}. Therefore, if the signal displays a right bend, then the value of $\beta$ becomes less relevant, whereas for a left bend it plays a more critical role. 
    \item $\bm{H_1}$: focusing now on the subplots depicting $\beta$ vs.\ $\log_{10}H_1$, it is possible to estimate the relative uncertainty on $H_1$. From figures~\ref{fig:curva_-3.4_gl}--\ref{fig:curva_-3.3_gl}, the relative uncertainties at  68\% CL are approximately 62\%, 135\%, 18\% and 63\%, respectively. 
    The variation in uncertainties can be explained by the fact that when the left bend lies far from the detection curve or the amplitude is low, the uncertainty $\Delta H_1$ increases, due to the behaviour shown in figure \ref{fig:variacion_parametros}. Therefore, the left bend (figure~\ref{fig:curva_-3.5_gl}) is characterized by a more precise estimation of $H_1$ compared to when one has a plateau or a right bend. In other words, there is a degeneracy of the effect of $H_1$ and $\beta$ unless one observes a left bend. 
    \item $\bm{m}$: 
    in the cases of a high-SNR plateau and a left bend (respectively, figure~\ref{fig:curva_-3.4_gl} and \ref{fig:curva_-3.5_gl}), at 95\% CL no upper bound smaller than the Planck mass can be set. The lower bound of $10^{-8}$ is due to the signal becoming undetectable below this value. Therefore, if LISA detected a plateau or a left-bend feature, no constraints could be placed on $m$. However, if the degeneracy was broken by choosing a value of $m$ that results in a right-bend feature in the LISA window (figure~\ref{fig:curva_-3.3_gl}), then the relative uncertainty would drop to $\Delta m/m\approx 18\%$ at 68\% CL. Figure~\ref{fig:curva_-3.9_gl} (low-SNR plateau) shows an upper bound on this parameter that occurs because the mean of the posteriors (red curve) fails to reconstruct accurately the position of the right bend in the injected signal (yellow curve) and places such bend within the LISA sensitivity curve, despite this not being the intended behaviour of the injected signal. Therefore, constraints of this type should be analyzed with extra care when the SNR is low.
    \item $\bm{\sigma_i}$: for a left-bend feature (figure~\ref{fig:curva_-3.5_gl}), $\sigma_i > 0.85$ at 68\% CL. If a high-SNR plateau or a right-bend feature were present (respectively, figure~\ref{fig:curva_-3.4_gl} and \ref{fig:curva_-3.3_gl}), then no constraints could be set, since $\sigma_i$ would not affect the signal amplitude within the explorable region. If a low-amplitude plateau was detected (figure~\ref{fig:curva_-3.9_gl}), then $\sigma_i$ would be minimally constrained: the likelihood distribution changes from roughly uniform over the entire prior range to having a higher probability for values below 0.4 and above 0.6. This type of bimodal distribution is typical of low-SNR signals.
\end{itemize}


\section{Conclusions}\label{sec:conclusions}


Regarding the four parameters that characterize the GWB in this model, the following conclusions can be drawn:

\begin{itemize}
    \item For the parameter $\beta$, the best constraints are obtained when the signal exhibits a left-bend feature, followed by the plateau shape for both high and low SNR cases. The poorest constraint is found in the right-bend scenario.
    
    \item Concerning $H_1$, the smallest relative uncertainty is achieved when the left-bend lies within the LISA sensitivity curve, in which case the relative uncertainty is $\Delta H_1/H_1 \approx 18\%$. However, if the left bend lies far from the sensitivity region, then the relative uncertainty can increase up to $63\%$. Furthermore, if the left bend lies far from the sensitivity region and the signal amplitude is low, then the relative uncertainty can exceed $100\%$, thus making this parameter indeterminable.
    
    \item For $m$, due to its degeneracy with other parameters, constraints can only be placed when the right bend lies within the sensitivity window, reaching a relative uncertainty of about $\Delta m /m \approx 18\%$ at 68\% CL.
    
    \item Regarding $\sigma_i$, a similar situation occurs but in the opposite direction: it can only be constrained when the left bend falls within the sensitivity range, reaching a relative uncertainty of approximately $\Delta \sigma_i / \sigma_i \approx 12\%$ at 68\% CL.
\end{itemize}

These results are physically meaningful, as both $H_1$ and $\sigma_i$ govern the behaviour of the left bend and meaningful constraints are obtained when the signal displays such a feature. The same applies to $\beta$, which determines the position of the left bend and, additionally, the slope of the curve. $m$ follows a similar trend but for the right bend.

To conclude, based on the analysis carried out, we find that LISA will be capable of placing significant constraints on the parameter space of the considered model, especially for the exponent $\beta$; the relative error on $H_1$, $m$ and $\sigma_i$ depends on the injected signal.

Should a power-law or a broken-power-law signal be observed, the pre-big-bang scenario would emerge as an appealing candidate to explain the data. This alone would not confirm the model itself, since there may be other early-universe models producing a similar signal at those frequencies and amplitudes. Nevertheless, in case this degeneracy were removed in favour of the pre-big-bang model (e.g., by polarization or multi-band observations spanning also cosmic-microwave-background scales), the implications for cosmology would be groundbreaking: not only would a detection suggest that general relativity requires an extension to the quantum realm but it would also provide a first empirical evidence supporting string theory, although it would not be a decisive one due to the theoretical limitations of the model described in section~\ref{sec:pbb_cosmology}.

Conversely, in the lack of detection all the parameters $H_1$, $m$, $\s_i$ and $\b$ would remain completely undetermined. This result, which we checked numerically, is somewhat obvious for $\s_i$ and $\b$ (since lower values of $0<\s_i<1$ only shift the left bend to lower frequencies and $\b$ only modifies the slope of the $\b=0$ plateau without affecting the global maximum value of the amplitude) but less so for $H_1$ and $m$. This inability to constrain the pre-big-bang model, if no GWB signal consistent with its predictions were observed within the LISA sensitivity range, is a typical case of ``over-parametrization,'' where the number of free parameters does not allow to extract predictions in the case of no detection. Such a case might motivate the development of alternative experimental strategies to overcome the limitations that prevented LISA from detecting the signal, for instance, by looking at multi-band measurements involving frequency ranges where the model has a foothold, such as in the band covered by the Einstein Telescope \cite{Ben-Dayan:2024aec}. 


\section*{Acknowledgments}

G.C.\ is supported by grant PID2020-118159GB-C41 funded by the Spanish Ministry of Science, Innovation and Universities MCIN/AEI/10.13039/501100011033. We warmly thank M.~Pieroni and J.~Torrado for invaluable discussions and support in the implementation, execution and problem-solving of the \texttt{SGWBinnner}. We also thank J.A.~Rubiño and C.~Perdomo for their helpful corrections and suggestions on an early version of this work as part of X.V.C.'s MSc thesis, as well as M.~Gasperini, E.~Pavone, R.~Rosati and, again, M.~Pieroni and J.~Torrado for detailed feedback on a first draft of the manuscript. This work was made possible also through the support of the WOST,  \href{https://withoutspacetime.org}{WithOut SpaceTime project}, supported by Grant ID 63683 from the John Templeton Foundation. The opinions expressed in this work are those of the authors and do not necessarily reflect the views of the John Templeton Foundation.

\appendix
\section{Derivation of \Eq{eq:def_freqs}}
\label{appA}

Let us start with $f_1$ \cite[eq.~(B.6)]{Ben-Dayan:2024aec}:
\ben
    f_1 = \frac{3.9\times10^{11}}{2\pi} \left( \frac{H_1}{\Mpl} \right)^{1/2}\left( \frac{z_\sigma}{z_d} \right)^{1/2},
\een
where here and in the following we omit the Hz units. To evaluate this expression, one must refer to eq.~(4.9) of the same paper to obtain the values of $z_\sigma$ and $z_d$:
\ban
z_\sigma &\simeq& \left( \frac{H_1}{\Mpl} \right)^{1/2} \left( \frac{m}{\Mpl} \right)^{-1/2} \left( \frac{\sigma_i}{\Mpl} \right)^{-2} \overset{\Mpl=1}{=} H_1^{1/2} \ m^{-1/2} \ \sigma_i^{-2}\,,\\
z_d &=& z_\sigma \ m^{-2/3} \ \sigma_i^{4/3} = H_1^{1/2} \ m^{(-4-3)/6} \ \sigma_i^{(-6+4)/3} = H_1^{1/2} \ m^{-7/6} \ \sigma_i ^{-2/3}\,.
\ean
Then,
\ben
\frac{z_\sigma}{z_d} = m^{(-3+7)/6} \ \sigma_i^{(-6+2)/3} =  m^{2/3}\ \sigma_i^{-4/3},
\een
so that
\begin{equation}
    f_1 = \frac{3.9\times10^{11}}{2\pi} H_1^{1/2} \ m^{1/3} \ \sigma_i^{-2/3}.
\end{equation}
Regarding $f_\sigma$, using once again \cite[eq.~(4.9)]{Ben-Dayan:2024aec}, we get $f_\sigma = f_1/z_\sigma$ and
\begin{equation}
f_\sigma 
= \frac{3.9\times10^{11}}{2\pi} m^{5/6} \ \sigma_i^{4/3}.
\end{equation}
Following the same process for $f_d$, we obtain $f_d = f_\sigma z_\sigma/z_d$ and
\begin{equation}
    f_d 
    = \frac{3.9\times10^{11}}{2\pi} m^{3/2} .
\end{equation}
Last, $f_s$ can be obtained from its definition \cite[eq.~(4.8g)]{Ben-Dayan:2024aec}:
\begin{equation}
f_s = \frac{f_1}{z_s} = \frac{3.9\times10^{11}}{2\pi z_s} H_1^{1/2} \ m^{1/3} \ \sigma_i^{-2/3}.
\end{equation}


\section{Derivation of \Eq{eq:limites}}
\label{appB}

In this appendix, we manipulate \Eqqs{eq:limitesf}--\Eq{eq:limites2} to extract the prior bounds on the fundamental theoretical parameters of the model. Let us begin by rewriting \Eq{eq:limites1} as a function of the parameters $H_1$, $m$ and $\sigma_i$ (in $\Mpl=1$ units):
\begin{equation*}
    \log_{10}H_1 + \frac{3}{2} \log_{10} \left( H_1^{1/2}m^{-7/6}\sigma_i^{-2/3} \right) - \frac{7}{2} \log_{10}\left( H_1^{1/2}m^{-1/2}\sigma_i^{-2} \right) \leq0 
\end{equation*}
Thus, the factors in $H_1$ and $m$ cancel out, yielding $\log_{10}\sigma_i \leq 0$ and
\begin{equation}\label{eq:cotasigma}
\sigma_i \leq 1\,.
\end{equation}
Next, we rewrite \Eq{eq:limites2}:
\begin{equation*}
    \log_{10}H_1 - 3 \log_{10} \left( H_1^{1/2} m^{-7/6} \sigma_i ^{-2/3} \right) + \log_{10} \left( H_1^{1/2} m^{-1/2} \sigma_i ^{-2} \right) > -42 - \log_{10}4 \approx -42.6
\end{equation*}
The factors of $H_1$ and $\sigma_i$ cancel out:
\begin{equation}\label{eq:cotam}
3\log_{10} m > -42.6 \qquad \Longrightarrow \qquad m >10^{-14.2}.
\end{equation}
From \Eq{eq:limitesf},
\begin{equation*}
f_\sigma > f_d \qquad \Longrightarrow \qquad \frac{3.9 \times 10^{11}}{2\pi} m^{5/6} \sigma_i^{4/3} > \frac{3.9 \times 10^{11}}{2\pi} m^{3/2},
\end{equation*}
and solving for $m$ we obtain
\begin{equation}\label{eq:cotafinm}
    m^{3/2 - 5/6} = m^{2/3} < \sigma_i ^{4/3} \qquad \Longrightarrow \qquad m < \sigma_i ^2 \,.
\end{equation}
Another condition from \Eq{eq:limitesf} is
\begin{equation*}
f_1 \gtrsim f_\sigma \qquad \Longrightarrow \qquad H_1^{1/2}m^{1/3}\sigma_i^{-2/3} \gtrsim m^{5/6}\sigma_i^{4/3}.
\end{equation*}
Solving for $H_1$,
\begin{equation}\label{eq:cotafinH}
 H_1 \gtrsim m \sigma_i^4\,,
\end{equation}
and by substituting the value obtained in \Eq{eq:cotasigma} into \Eq{eq:cotafinm} we have
\begin{equation}
    m < 1\,.
\end{equation}
The final upper bound for $H_1$ can be derived by recalling that the kinematic details of the bounce are expected to affect only the very-high-frequency modes that cross the horizon during the bounce phase itself. For further details, see~\cite{Gasperini:2016gre}. Hence,
\begin{equation}
    H_1<1\,.
\end{equation}

A similar procedure can be applied to the lower bounds by substituting the value from \Eq{eq:cotam} into \Eq{eq:cotafinm}:
\begin{equation}
    \sigma_i > 10^{-14.2 / 2}= 10^{-7.1}\,,
\end{equation}
and introducing both values into \Eq{eq:cotafinH} we get
\begin{equation}
H_1 \gtrsim 10^{-14.2} \times 10^{-7.1 \times 4} = 10^{-42.6}\,.
\end{equation}


\end{document}